\documentclass[conference]{IEEEtran}
\IEEEoverridecommandlockouts

\usepackage{cite}
\usepackage{amsmath,amssymb,amsfonts}
\usepackage{algorithmic}
\usepackage{graphicx}
\usepackage{textcomp}
\usepackage{xcolor}
\usepackage{multirow}
\usepackage{color}
\usepackage[T1]{fontenc}
\usepackage{pifont}
\newcommand{\cmark}{\textcolor{green!80!black}{\ding{51}}}
\newcommand{\xmark}{\textcolor{red}{\ding{55}}}
\usepackage{makecell}

\def\BibTeX{{\rm B\kern-.05em{\sc i\kern-.025em b}\kern-.08em
    T\kern-.1667em\lower.7ex\hbox{E}\kern-.125emX}}
\begin{document}

\title{Mobile Interaction for Assessing Fatigue, Sleep, and Activity in Neurodegenerative and Chronic Diseases}

\author{

\IEEEauthorblockN{ Julian Fierrez\IEEEauthorrefmark{1}, Alejandro Peña\IEEEauthorrefmark{1}, Aythami Morales\IEEEauthorrefmark{1}, Ruben Tolosana\IEEEauthorrefmark{1}, Ruben Vera-Rodriguez\IEEEauthorrefmark{1}, }

\IEEEauthorblockN{ Meenakshi Chatterjee\IEEEauthorrefmark{2}, Ahmaniemi Teemu\IEEEauthorrefmark{3}, Wan-Fai Ng\IEEEauthorrefmark{4}, Walter Maetzler\IEEEauthorrefmark{5}, Nikolay V. Manyakov\IEEEauthorrefmark{6}, }

\IEEEauthorblockN{ Jennifer Kudelka\IEEEauthorrefmark{7}, Ralf Reilmann\IEEEauthorrefmark{8}, C. Janneke van der Woude\IEEEauthorrefmark{9}, Kristen Davies\IEEEauthorrefmark{4}, Victoria Macrae\IEEEauthorrefmark{4}, }

\IEEEauthorblockN{ IDEA-FAST Consortium }

\vspace{.2cm}

\IEEEauthorblockA{\IEEEauthorrefmark{1}BiometricsAI, Universidad Autonoma de Madrid (UAM), Spain}

\IEEEauthorblockA{\IEEEauthorrefmark{2}Johnson \& Johnson, Spring House, PA, USA}

\IEEEauthorblockA{\IEEEauthorrefmark{3}VTT Technical Research Centre of Finland, Finland}

\IEEEauthorblockA{\IEEEauthorrefmark{4}Newcastle upon Tyne Hospitals NHS Foundation Trust, Newcastle upon Tyne, UK}

\IEEEauthorblockA{\IEEEauthorrefmark{5}Dept. of Neurology, University Hospital Schleswig-Holstein and Kiel University, Kiel, Germany}

\IEEEauthorblockA{\IEEEauthorrefmark{6}Johnson \& Johnson, Beerse, Belgium}

\IEEEauthorblockA{\IEEEauthorrefmark{7}Dept. of Neurology of the University Hospital in Kiel, Germany}

\IEEEauthorblockA{\IEEEauthorrefmark{8}George Huntington Institute, Münster, Germany}

\IEEEauthorblockA{\IEEEauthorrefmark{9}Dept. of Gastroenterology and Hepatology, Erasmus University Medical Centre, Rotterdam, The Netherlands}

}


\maketitle

\begin{abstract}
Fatigue, sleep, or disturbances in daily activities are common symptoms among patients with neurodegenerative disorders (NDD) and immune-mediated inflammatory diseases (IMID). The current assessment of such symptoms is usually conducted using patient reported outcomes (PROs) based on standardized questionnaires that patients usually complete every few months. This assessment protocol has raised some concerns, due to its propensity to exhibit biases derived from its subjectivity nature, or the low sensibility to changes, which may lead to a failure when trying to capture variability over time. In this work, we explore the use of smartphone data, which can serve as a proxy for how patients interact with their devices, to provide an effective, reliable, and objective assessment of the symptoms mentioned above. Our study comprises data from $137$ participants belonging to $6$ different disease groups, plus a healthy control group. We conducted statistical analysis based on repeated measures correlation, in which we analyze the correlation between screen-time and app-usage features with scores obtained from the PROs collected from the participants using a smartphone application.
\end{abstract} 


\section{Introduction}\label{sec:intro}

Patients suffering from chronic diseases such as neurodegenerative disorders (NDD) and immune-mediated inflammatory diseases (IMID) experience a significant reduction in their quality of life as a result of these conditions. For example, both Parkinson's Disease patients (estimated at around $1.8\%$ of people over $60-70$ years of age~\cite{de2000prevalence}) and Huntington's Disease patients usually suffer from health issues such as rigidity, speech disturbances, or tremors, which ultimately affect their daily living. Among the most commonly reported symptoms shared by patients with NDD and IMID, we find fatigue and sleep disturbances~\cite{arends2017physical,chavarria2019prevalence,hewlett2011fatigue,morgan2018lupus}. These symptoms not only affect their daily activities, they are also among the first symptoms of patients in the early stages of the disease~\cite{poewe2017parkinson}.

Despite the relevant impact of fatigue and sleep disturbances on daily activities of patients, current information on these symptoms  consists of feedback directly from patients, called Patient Reported Outcomes (PROs). PROs are commonly used in IMID and NDD treatments to complement other clinical endpoints. The lack of agreement on how to efficiently measure these health issues, along with the inner subjectivity of the methods currently proposed, complicate their assessment. Evaluation of such symptoms is usually conducted using generic questionnaires~\cite{hewlett2011measures}, which are often filled out during clinical visits. Due to their relevance, disease-specific questionnaires are also used for the evaluation of fatigue and sleep, where we can cite the PDSS, PDSS-2 and SCOPA-scale~\cite{kurtis2018review} for Parkinson's Disease, LupusQoL~\cite{mcelhone2007lupusqol} for Systematic Lupus Erythematosus, or the IBD-F patient self-assessment scale~\cite{czuber2014ibd} in the case of Inflamatory Bowel Disease, among others. However, since long periods of time may elapse between the completion of the questionnaires (e.g. questionnaires completed during clinical visits), these measures may fail to capture fine variability of the assessed symptoms, whose intensity varies from day to day.

Therefore, there is an urge for new ways to assess patients with NDD and IMID, as well as the inclusion of new measures of fatigue and sleep, with information at least comparable to that of PROs, as reliable endpoints of treatments. A key factor for any successful intervention of patients is improving their quality of life related to health. Thus, a more continuous follow-up would allow for a more adequate treatment, ultimately improving their condition. In recent years, some advances have been made towards more efficient and continuous clinical treatments, which include patients' health status assessment via digital measures. The use of mobile sensors and data \cite{2020_CDS_HCIsmart_Acien,acien2021becaptcha,2022_IJCB_MobileB2C_Straga} to collect clinical information \cite{Blanca25AI4Food} has been studied in the context of Parkinson's Disease interventions~\cite{tripathi2022clinical,hu2020}, and is starting to be very successful in many other applications of e-health that involve continuous monitoring \cite{2023_Database_AI4FoodDB_Romero}. Regarding NDD, for example, the authors of~\cite{orozco2020apkinson} developed a mobile application to improve the monitoring of Parkinson's patients progression based on gait in walking, speech pronunciation and hand movements. Other biometric modalities shown to be useful in monitoring NDDs include face dynamics \cite{2021_CVPRW_Parkinson_Gomez,2023_PLOS_FacialParkinson_Gomez}, keystroking \cite{Tiago25CHI,Acien25keys}, and handwriting \cite{2019_FG_PDhandw_Castrillon,2020_COGN_HandwTrends_Faundez}. All of these technologies can be framed in what is known as mobile health (mHealth)~\cite{rehg2017mobile}. Note that the benefits of using digital measures go beyond continuous monitoring; e.g., they can also enable adequate individual personalization \cite{Sergio26personal}.

Recently, the IDEA-FAST project has reunited researchers from different universities and health organizations with the aim of identifying digital endpoints that could serve as a reliable assessment of patients with NDD and IMID \cite{idea-fast23concept}. To this end, different digital measures and their potential relation with PROs of fatigue, sleep, and daily activities have been studied in two phases: Feasibility Study (FS) \cite{Pinaud24IDEA-FS} and Clinical Observational Study (COS) \cite{Walter26IDEA-COS}. In the present paper, we present initial research within the Feasibility Study exploring the use of mobile interaction data as an objective measure of fatigue, sleep, and activity in patients with NDD and IMID. The purpose of the work is to analyze whether the collected data on the mobile device can be as informative as the feedback reported within standard PRO questionnaires. Our study comprises $137$ participants from $6$ different disease groups, namely Parkinson's Disease, Huntington's Disease, Rheumatoid Arthritis, Systematic Lupus Erythematosus, Primary Sjogren's Syndrome, and inflammation of the digestive tract, plus a healthy control group. We use a smartphone application to collect data that reflect how participants interact with their devices. The application is also used to obtain periodic PROs from the participants using mobile prompted questionnaires related to fatigue, sleep, and daily activities of the participants. We extract hand-made features from mobile data and analyze their association with PROs based on repeated measures correlation~\cite{bakdash2016repeated} to assess whether mobile interaction patterns could be potentially valid measures for predicting symptoms studied in this work.


\section{Methodology}\label{sec:methodology}


\subsection{Data Collection}\label{sec:data_collection}

\begin{figure}
\includegraphics[width=\columnwidth]{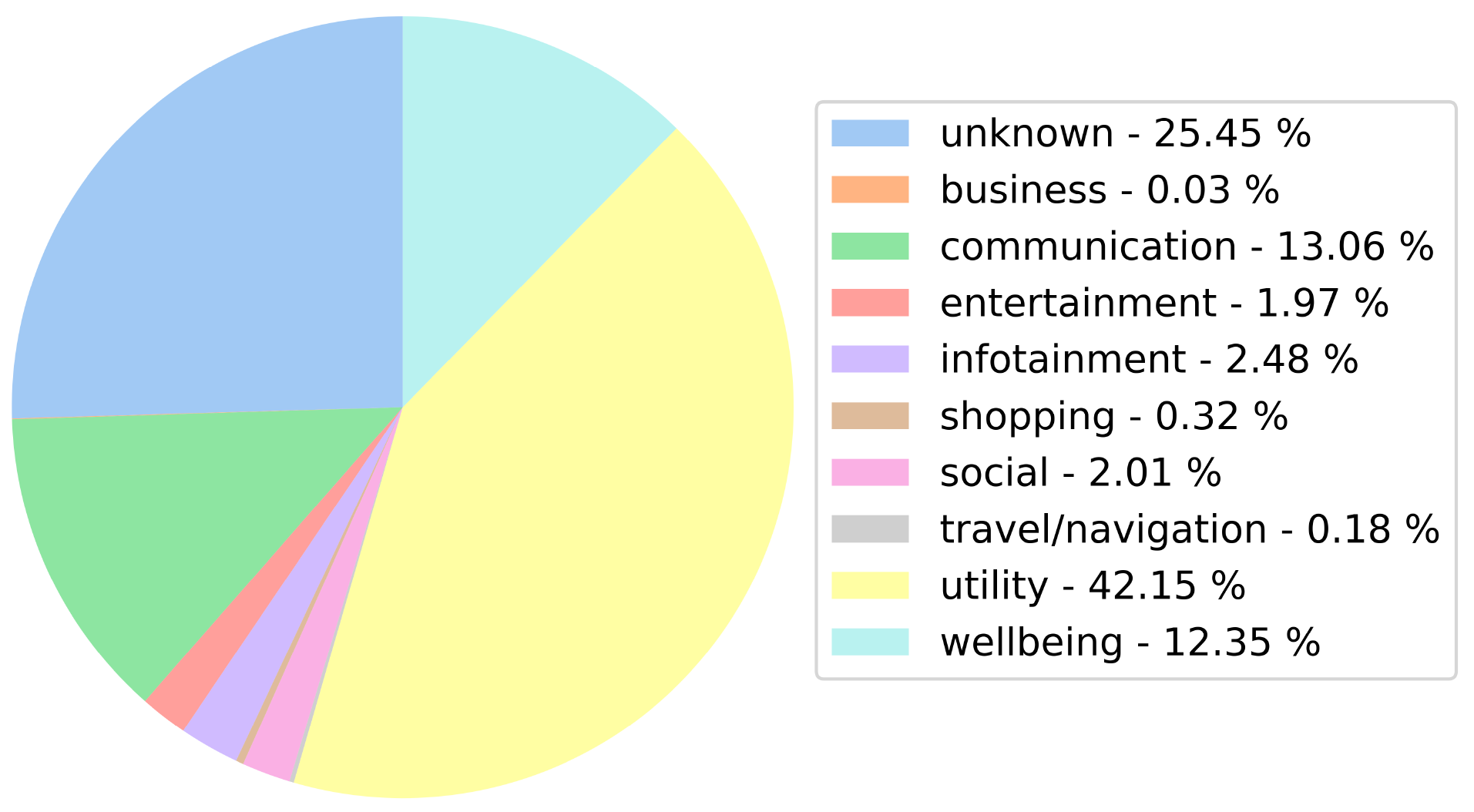}
\caption{App categories distribution in the clean files.} \label{fig:app_distribution}
\end{figure}

Data used in this study were collected using a mobile phone application developed as part of the IDEA-FAST Feasibility Study (FS). The application registered data reflecting how the participants interact with their devices. Each data file contains mobile records including a time stamp, a data type, a data value, and a geo-location code. Different data types can be found in the records, as the application registered a variety of events in the devices. The main data types are as follows:
\begin{itemize}
    \item PHONE\_APP\_IN\_FOREGROUND: Applications running in the foreground. These include an application code as the data value in which the app category can be found. The applications were grouped into $11$ different categories (see Figure~\ref{fig:app_distribution}) by the application.
    \item PHONE\_SCREEN: Screen on/off events.
    \item PHONE\_BATTERY: Battery level updates.
    \item PHONE\_ACTIVITY: Record reflecting the participant's current activity as discrete states (e.g., whether the participant is still, walking, on bike, etc.)
    \item QUESTIONNAIRE: Records generated when the participant filled out a questionnaire for the study.
\end{itemize}

Note that these records were generated whenever a related event occurred in the mobile phones (e.g. the participant filled a questionnaire, opened an application, etc.). Hence, the mobile collected data do not have a fixed sample frequency.

Data for the FS study were collected from a group of 137 different participants from $6$ individual disease groups, including patients with Parkinson's Disease (PD $= 18$), Huntington's Disease (HD $= 9$), Rheumatoid Arthritis (RA $= 17$), Systematic Lupus Erythematosus (SLE $= 17$), Primary Sjogren's Syndrome (PSS $= 17$), Inflammatory Bowel Disease (IBD $= 17$), plus a healthy control group (HC $= 42$). Participants were selected and assigned to four clinical sites (i.e., Kiel $26.95\%$, Newcastle $40.43\%$, EMC Rotterdam $19.85\%$, and GHI Muenster $12.77\%$), where demographic information and medical data were collected at the beginning of the study. The study is composed of $35.46\%$ men and $64.54\%$ women. The mean age of the participants is nearly $52$ years, with $9.2\%$ under $30$ years, $20.56\%$ between $30$-$40$ years, $12.05\%$ between $40$-$50$ years, $21.98\%$ between $50$-$60$ years, $19.14\%$ between $60$-$70$ years, and the remaining $17.02\%$ over $70$ years. Among all participants, $10$ installed the mobile application on their private phones, while others used it on study-provided phones.

\begin{table*}[t]
    \centering
    \small
    \begin{tabular}{l|l|c|c|c|c}
  
        \hline
        \multirow{2}{*}{\textbf{ID}} & \multirow{2}{*}{\textbf{Question}} &\multicolumn{4}{c}{\textbf{Questionnaire}}\\
        \cline{3-6}
        &&Morning&\makecell{Early\\afternoon}&\makecell{Late\\afternoon}&Evening\\
        \hline\hline
        \textit{Feel\_Q$1$} & Physical fatigue &\cmark&\cmark&\cmark&\cmark\\
        \hline
        \textit{Feel\_Q$2$} & Mental fatigue &\cmark&\cmark&\cmark&\cmark\\
        \hline
        \textit{Feel\_Q$3$} & Anxiousness &\cmark&\cmark&\cmark&\cmark\\
        \hline
        \textit{Feel\_Q$4$} & Depression &\cmark&\cmark&\cmark&\cmark\\
        \hline
        \textit{Feel\_Q$5$} & Pain &\cmark&\cmark&\cmark&\cmark\\
        \hline\hline
        \textit{To\_Bed\_Time} & I went to bed at &\cmark&\xmark&\xmark&\xmark\\
        \hline
        \textit{Woke\_Up\_Time} & I woke up &\cmark&\xmark&\xmark&\xmark\\
        \hline
        \textit{Sleep\_Details\_Q$1$} & How was your sleep? &\cmark&\xmark&\xmark&\xmark\\
        \hline
        \textit{Sleep\_Details\_Q$2$} & Time to fall asleep &\cmark&\xmark&\xmark&\xmark\\
        \hline
        \textit{Sleep\_Details\_Q$3$} & Time awake during night &\cmark&\xmark&\xmark&\xmark\\
        \hline
        \textit{Sleepiness} & Sleepiness, current feeling &\xmark&\cmark&\cmark&\cmark\\
        \hline\hline
        \textit{Activities\_Q$1$} & \makecell{My activities of the day,\\ physically} &\xmark&\xmark&\xmark&\cmark\\
        \hline
        \textit{Activities\_Q$2$} & \makecell{My activities of the day,\\ mentally} &\xmark&\xmark&\xmark&\cmark\\
        \hline
        \textit{Activities\_Q$3$} & Other comments &\xmark&\xmark&\xmark&\cmark\\
        \hline
        
    \end{tabular}
\caption{Patient reported outcomes (PRO) prompted in the questionnaires.}
\label{tab:pros}
\end{table*}
\normalsize

Patient reported outcomes (PROs) were collected using the mobile application, which prompted questionnaires $4$ times per day. The questionnaires were asked at $9$:$00$, $13$:$00$, $17$:$00$, and $21$:$00$, each with a different number of questions. Each questionnaire was available for $3$ hours, except for the evening questionnaire, which was available for $2$ and a half hours. The application recorded whether the participant responded to each question and the response time. Table~\ref{tab:pros} presents the different PROs collected and in which daily questionnaires they appeared. Note that we can group the questions into PROs related to fatigue, sleep, and daily activities (i.e., the three main symptoms analyzed in this work). The majority of the questions accepted a response on the Likert scale (with a predefined scale of $7$ levels, ranging from $0$ to $6$). Both \textit{To\_Bed\_Time} and \textit{Woke\_Up\_Time} accepted clock-type responses, while sleep-related questions had a drop-down menu with different options, except for \textit{Sleep\_Details\_Q}$\mathit{1}$, which also had a Likert scale format. The final question, \textit{Activities\_Q}$\mathit{3}$, accepted a free text response. Such a format has little interest with regard to the scope of this work, so this last question will not be considered in the following sections.

Together with the daily collection of mobile PRO, the participants conducted clinical visits, where they completed traditional PRO surveys. These surveys included the Functional Assessment of Chronic Illness Therapy Fatigue Scale (FACIT-F) and the Medical Outcomes Study Sleep Scale (MOS-SS) acute. The FACIT-F is a fatigue assessment questionnaire, while the MOS-SS acute focus is on sleep disturbances. Participants completed these questionnaires once a week, after every period of device use of other sensors and digital measures studied within the IDEA-FAST FS. The purpose of capturing traditional PROs together with digital measures was to assess the equivalence of both traditional and digital measures.

\subsection{Data Analysis}\label{sec:data_analysis}

As we exposed in the previous section, mobile data captured in the study comprise data records reflecting different interactions. After an initial inspection of the data, we decided to discard the records belonging to PHONE\_BATTERY or PHONE\_ACTIVITY data types in our work. Although the PHONE\_BATTERY records clearly do not provide significant information on the problem in question, the PHONE\_ACTIVITY records might first appear to be of some interest. However, the information in these is at a high level of abstraction compared to the one obtained with other sensors of the project, so we discarded activity as a variable of interest in the mobile-collected data analysis. Thus, we will focus ts on two variables: screen time and app usage (i.e., variables related to PHONE\_SCREEN and PHONE\_APP\_IN\_FOREGROUND records), which have been shown to be highly distinctive in related works \cite{2019_IBPRIA_Acien_Keystroke}, especially when combined with stronger biometrics \cite{2019_MULEA_Acien_MultiLock}. The experiments will be based on the data and time stamps of these records.

We aggregated the data into fixed-duration time windows and extracted hand-made features for the two variables of interest in each of them. For app usage, we use as features the number of appearances of each app category within the time window. This led us to $11$ different features, one for each app category. On the other hand, the screen time variable analysis is based on screen events. We define a SCREEN event as the interval between the closest consecutive SCREEN\_ON (start) and SCREEN\_OFF (end) PHONE\_SCREEN records (which may have other types of records between them). During the extraction of screen events from the data, we remove unmatched PHONE\_SCREEN records, that is, records that do not follow the SCREEN sequence SCREEN\_ON/SCREEN\_OFF. For example, when we encounter two consecutive SCREEN\_ON records, we remove the first one, as it has no SCREEN\_OFF pair to compute the screen time. Once we have the screen events, we extract as features in each data aggregation window statistical values such as the mean, median, standard deviation, and total screen time, as well as the number of events, and both the maximum and minimum screen time events.

Our experimental setup starts from a data quality analysis, where we filter good quality records. We then extract the previously introduced features and perform an association analysis based on repeated measures correlation~\cite{bakdash2016repeated}. The repeated measures correlation is a statistical approach to determine the linear association between two paired values assessed more than once in multiple subjects. Traditional correlation approaches such as Pearson or Spearman correlation assume the independence of the samples~\cite{howell2012statistical}, which can only be satisfied when each subject represents one data point. This leads to a violation of this principle when multiple data points are available for each subject, requiring an aggregation of the data points which can mislead the results~\cite{myung2000toward}. Repeated-means correlation, on the other hand, respects the principle of independence when multiple data points are available, making it a perfectly suited approach for our problem.

\section{Experiments and Results}\label{sec:experiments}

In this Section we will introduce the experiments carried out in this work, and the main results. Firstly, we analyze the quality of the collected data (Section~\ref{sec:preprocessing}), and then extract the features and coverage of the questionnaire from the cleaned data (Section~\ref{sec:aggregation}). Once we have extracted mobile features, we analyze the association between mobile features and PROs based on the correlation of repeated measures~\cite{bakdash2016repeated} (Section~\ref{sec:correlation}), as we introduced in Section~\ref{sec:data_analysis}. Finally, we repeat the association analysis by cohort group (Section~\ref{sec:cohort}), with the objective of exploring cohort-based effects.

\subsection{Data Cleaning and Preprocessing}\label{sec:preprocessing}

The data collected for the analysis comprise $150$ files from $137$ participants. Among the $150$ files, we found $9$ records with no data collected. In addition, there were $11$ records with data captured, but in which the participant did not provide any responses to the questionnaires. We discarded these $20$ records.

We then proceeded to analyze the number of days recorded in the remaining $130$ files. We found that most files have between $10$ and $35$ days recorded, with a mean of $23.4$ days and a median of $26$. Only $13$ records have less than $10$ days of data recorded, with $3$ of them presenting less than $3$ days. We also discarded these three records due to the few data captured. We also observed that some records present gaps in the data (i.e. periods of time when no data were captured, due to application breaks, for example). These gaps range from a few hours to even a week in some extreme cases. However, we did not find the gaps to be problematic as long as there was enough data in the records, so we did not discard any file.

Finally, we validated the time stamps of the good quality records \cite{2012_QualityBio_FAlonso} to ensure that the order of the samples is consistent with their time stamps. After the cleaning process, $127$ files were available from $118$ different participants.

\subsection{Data Aggregation, Feature Extraction, and Questionnaires}\label{sec:aggregation}

After the data cleaning process, we were ready to extract features from the records to associate the mobile data with the PROs. The feature extraction, as previously mentioned in Section~\ref{sec:data_analysis}, focused on screen events and app usage related records. We decided to use aggregates of $24$-hours (i.e., daily) for the data. We extracted features from the records available in each daily window and computed the PROs as the daily average of the responses to each question in the questionnaires. 
We also explored the use of shorter ($8$- and $12$-hours) and longer ($48$-hours) windows. Using shorter windows led to a significant reduction in the number of events in each period due to the low frequency of the mobile data, which ultimately affected the variability of the features extracted. Recall that both screen events and app usage data were recorded whenever related changes occur on the devices, and not at a high fixed frequency as other mobile measures such as accelerometer or touch dynamics data \cite{fierrez2018benchmarking,2020_IJDAR_ExploitCXsign_Tolosana}. Furthermore, because the participants did not fill all the questionnaires (or the questions in each questionnaire, see Table~\ref{tab:coverage}), daily averaging of the PROs helped us compensate for the gaps in the responses to the questions that appeared multiple times per day (see Table~\ref{tab:pros}). However, we did not find significant differences in the results when using longer windows.

We extracted app usage features as the number of events associated with each of the defined $11$ app categories. The value of each app record is one or more app codes, which are comma separated valued, where the left number indicates the category. Figure~\ref{fig:app_distribution} presents the distribution of the categories of the app in the clean $127$ files. As we can observe, there are huge differences in the app distribution. Roughly $70$\% of the applications belong to ``Unknown'' or ``Utility'', with ``Communication'' ($13.06$\%) and ``Wellness'' ($12.35$\%) being the categories with the most presence. The rest of the categories are almost absent from our data. Furthermore, only the ``Unknown'' and ``Utility'' categories are well distributed among the participants and data aggregation windows, with the other applications showing an irregular presence.

Taking into account the features of screen events, the statistical values for screen time could be computed in almost all cases. As we are now dealing with statistical values from the screen activations, we do not have the same problem as with the app features (i.e., some app category occurrences were zero for a significant number of data windows and participants). Here, we discarded screen events with a duration greater than $3$ hours. We selected this threshold after inspecting that, over such a limit, screen events take an extremely long duration. We found in different cases that these longer screen time events were caused by application breaks and gaps in the data, so we consider these events as outliers. 

\begin{table*}[]
    \centering
    \begin{tabular}{c|c|c|c|c|c|c|c|c|c|c|c|c}
    \hline
     F$1$&F$2$&F$3$&F$4$&F$5$&B\_Time&W\_Time&S$1$&S$2$&S$3$&Sleep&A$1$&A$2$\\
     \hline\hline
     $0.868$&$0.854$&$0.440$&$0.369$&$0.604$&$0.741$&$0.751$&$0.622$&$0.763$&$0.763$&$0.923$&$0.381$&$0.382$\\
     \hline
     
    \end{tabular}
    \caption{Coverage of the questionnaire for each PRO measured in daily average aggregates. See Table~\ref{tab:pros} for the meaning of each PRO.}
    \label{tab:coverage}
\end{table*}

Finally, we would like to extract the questionnaire coverage, i.e. the percentage of responses available for each question. As we commented before, we will use daily aggregates; therefore, PROs are represented by the daily average of responses. Therefore, the coverage of the questionnaire represents the percentage of days with a valid response during the data collection period. Table~\ref{tab:coverage} presents the coverage results. We can notice significant differences between some PROs. Although the coverage is high for the \textit{Feel\_Q}$\mathit{1}$ and \textit{Feel\_Q}$\mathit{2}$, both associated with fatigue, it is significantly decreased for the \textit{Feel\_Q}$\mathit{3}$ and \textit{Feel\_Q}$\mathit{4}$ questions. This difference is quite surprising, as these questions were prompted together in all the questionnaires of the day (see Table~\ref{tab:pros}). The sleep related questions and \textit{To \_Bed \_Time} and \textit{Woke \_Up \_Time} present a coverage over $60\%$, with the \textit{Sleepiness} question having the highest coverage with $92.3\%$. On the other hand, the coverage of the daily activities questions does not exceed $40\%$. These questions were only asked in the evening questionnaire, which leads us to believe that this questionnaire was the one with the least number of responses.

\subsection{Association Analysis}\label{sec:correlation}

As we mentioned earlier, the analysis of the potential associations between mobile features and PROs will be based on the correlation of repeated measures. After the cleaning process of Section~\ref{sec:preprocessing}, a total of $2957$ and $2858$ data points of $118$ participants were available for screen events and application usage variables, respectively. As we are using daily aggregates for the data, each data point corresponds to a day in which at least one feature could be calculated correctly. This is the reason for the difference in data points between the two variables studied here. During the repeated measures correlation computation, we only considered feature-PRO pairs where a PRO score was available. Recall that not all participants filled all questionnaires during the data collection period and even not all questions in each questionnaire were necessarily answered (see Table~\ref{tab:coverage}). Thus, each feature-PRO pair has a specific number of data points. 

Firstly, we will analyze the association of the app usage features with the PROs. Here, we encountered a problem when computing the correlation values. The few appearances of some categories (see Figure~\ref{fig:app_distribution}) restricted the number of available data points. However, some categories with a significant presence, for example ``Communication'' or ``Health'', failed when trying to compute correlation values. We inspected such categories in the participants data, finding that in most cases they have a high presence only in a few days while very scarce in the rest. We hypothesized that their irregular presence could be the reason for the error. These considerations restricted the association analysis to just two categories, ``Unknown'' and ``Utility'' apps. The correlation coefficients of repeated measures $r_m$, along with the statistical significance $\rho$ and the number of data points, are presented in Table~\ref{tab:app_correlation}. The first thing we can notice is the huge difference between the number of points available for each feature/PRO pair. This difference is directly related to the coverage of the questionnaire that we saw in Table~\ref{tab:coverage}. Taking into account the values of $r_{m}$, we cannot find a significant correlation value between any feature-PRO pair. We find the strongest $r_{m}$ value between the ``Unknown'' category and the \textit{Activities\_Q}$\mathit{2}$ PRO (with $\rho < 0.01$). In addition to the weak correlation values, most pairs show a large significance value. This result, along with the restriction in the computation of most of the app categories, leads us to discard the app usage variable in a further detailed analysis.

\begin{table}[]
    \centering
    \tiny
    \begin{tabular}{l|l|c|c|c|c|c}
    \hline
    \multirow{2}{*}{\textbf{PRO}}&\multirow{2}{*}{\textbf{Question}}&\multirow{2}{*}{\textbf{\# points}}&\multicolumn{2}{c|}{\textbf{Unknown}}&\multicolumn{2}{c}{\textbf{Utility}}\\
    \cline{4-7}
    &&&$\mathbf{r_m}$&$\mathbf{\rho}$&$\mathbf{r_m}$&$\mathbf{\rho}$\\
    \hline\hline
    Feel\_Q$1$&Physical fatigue&$1979$&$0.026$&$0.255$&$0.038$&$0.103$\\
    \hline
    Feel\_Q$2$&Mental fatigue&$1948$&$0.075$&$0.001$&$0.067$&$0.004$\\
    \hline
    Feel\_Q$3$&Anxiousness&$1005$&$0.053$&$0.108$&$0.057$&$0.088$\\
    \hline
    Feel\_Q$4$&Depression&$842$&$0.018$&$0.617$&$-$&$-$\\
    \hline
    Feel\_Q$5$&Pain&$1376$&$-0.050$&$0.070$&$-0.031$&$0.267$\\
    \hline\hline
    To\_Bed\_Time&I went to bed at&$1691$&$-0.030$&$0.234$&$-0.019$&$0.452$\\
    \hline
    Woke\_Up\_Time&I woke up&$1712$&$-0.024$&$0.333$&$-0.059$&$0.019$\\
    \hline
    Sleep\_Details\_Q$1$&How was your sleep?&$1416$&$-0.079$&$0.004$&$-0.065$&$0.019$\\
    \hline
    Sleep\_Details\_Q$2$&Time to fall sleep&$1741$&$0.036$&$0.153$&$0.022$&$0.375$\\
    \hline
    Sleep\_Details\_Q$3$&Time awake at night&$1741$&$0.004$&$0.886$&$0.019$&$0.446$\\
    \hline
    Sleepiness&Sleepiness feeling&$2107$&$-0.027$&$0.230$&$-0.031$&$0.169$\\
    \hline\hline
    Activities\_Q$1$&\makecell{Day activities,\\ physically}&$872$&$0.021$&$0.556$&$0.007$&$0.851$\\
    \hline
    Activities\_Q$2$&\makecell{Day activities,\\ mentally}&$873$&$0.111$&$0.002$&$0.039$&$0.275$\\
    \hline
    \end{tabular}
    \caption{Repeated measures correlation coefficients $r_m$ and significance values $\rho$ between app and PROs.}
    \label{tab:app_correlation}
\end{table}
\normalsize

\begin{figure}
    \centering
    \includegraphics[width=\columnwidth]{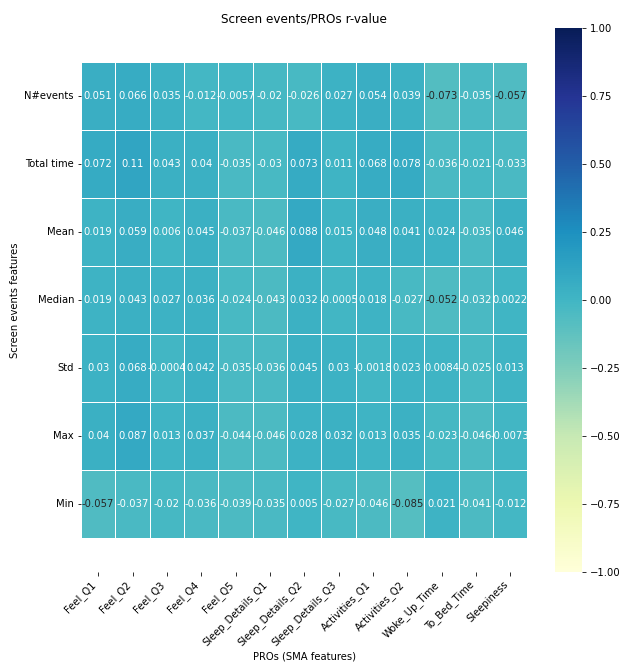}
    \caption{Repeated measures correlation coefficients $r_m$ computed between the screen events features and the PROs.}
    \label{fig:screentime_correlation}
\end{figure}

Now that we have analyzed the association of the PROs with the app usage variable, we will focus on the screen events. Figure~\ref{fig:screentime_correlation} presents the $r_m$ values computed between the screen event features and the PROs. The feature-PRO pair with the lowest number of data points is the ``Maximum screen time event''-\textit{Feel\_Q4} pair, with only $844$ data points. On the other hand, the pair ``Minimum screen time event''-\textit{Sleepiness} has $2124$ points, being the one with the most coverage. We can observe in Figure~\ref{fig:screentime_correlation} a situation similar to the one we saw in the case of app usage. The results of our analysis do not show a significant linear correlation between any pair. We can find the strongest value $r_m$ in the pair ``Total screen time''-\textit{Feel\_Q2} ($r_m = 0.11, p < 0.01 $), which is not large enough to denote a clear linear correlation.

\subsection{Cohort-based Association Analysis}\label{sec:cohort}

The results of Section~\ref{sec:correlation} showed slight correlation values between the features extracted from the captured mobile data (i.e., the usage characteristics of the app and the events on the screen) and the PROs, which ultimately seems to indicate that no clear linear correlation can be established between them. However, we conducted such an analysis on all participants and therefore specific cohort-related patterns may have been minimized. In this Section, we analyze the association by cohort group of the PROs with the screen events features only, as we commented in previous analysis the problems associated with the app usage features.

Table~\ref{tab:cohort_correlation} presents the highest associations between pairs calculated by cohort group. Note that, as we are computing the correlation values using only the data of each cohort, the number of data points available have significantly decreased. We can observe values $r_m$ much stronger than those in Figure~\ref{fig:screentime_correlation}, which confirms our hypothesis (i.e., computing the association between all participants masks cohort-specific patterns). Now, we find $4$ associations with a $|r_m|$ value over $0.3$ in the Huntington Disease group. In fact, the strongest association is a negative correlation of $-0.444$ between the number of screen events and the \textit{To\_Bed\_Time} PRO. Note that two of these associations are pairs that include the \textit{Feel\_Q}$\mathit{2}$ PRO (i.e. mental fatigue), with the other two being related to the time when the participant went to bed and the time to fall asleep, which are closely related. Other cohorts, such as Rheumatoid Arthritis, Systematic Lupus Erythematosus or Primary Sjogren's Syndrome, present stronger associations with different PROs. Nevertheless, none of the values in Table~\ref{tab:cohort_correlation} is higher than $0.5$, so we cannot conclude that a linear correlation can be established between any of these pairs. Furthermore, finding stronger values in this analysis could be explained by the existence of some cohort specific patterns, but the low number of data points could be affecting the results, so it would be good to assess further the results.

\begin{table}[]
    \centering
    \tiny
    \begin{tabular}{l|c|c|c|c|c}
    \hline
    \textbf{Feature/PRO pair}&\textbf{Cohort}&$\mathbf{r_m}$&$\mathbf{\rho}$&\textbf{\# points}&\textbf{\# part.}\\
    \hline\hline
    \# events $-$ To\_Bed\_Time & HD &$-0.444$&$<0.01$ &$61$&$8$\\
    \hline
    Screen time std $-$ Feel\_Q$2$&HD&$0.416$&$<0.01$&$76$&$8$\\
    \hline
    Total screen time $-$ Activities\_Q$1$&RA&$0.354$&$0.05$&$44$&$15$\\
    \hline
    Mean screen time $-$ Sleep\_Details\_Q$2$&HD&$0.331$&$0.014$&$64$&$8$\\
    \hline
    Maximum screen time $-$ Feel\_Q$2$&HD&$0.327$&$<0.01$&$75$&$8$\\
    \hline
    Median screen time $-$ Activities\_Q$2$&SLE&$-0.306$&$0.02$&$70$&$13$\\
    \hline
    Total screen time $-$ Feel\_Q$4$&PSS&$0.305$&$<0.01$&$130$&$15$\\
    \hline
    Mean screen time $-$ Feel\_Q$3$&RA&$-0.304$&$<0.01$&$76$&$15$\\
    \hline
    \# events $-$ Activities\_Q$2$&PD&$0.3$&$0.02$&$66$&$16$\\
    \hline
    
    \end{tabular}
    \caption{Top screen event feature $-$ PRO pairs by cohort group.}
    \label{tab:cohort_correlation}
\end{table}
\normalsize

After assessing the associations by cohort group, we would like to check whether different patterns can occur equally across different demographic groups. More specifically, we conducted the association analysis using the gender and age information available from the participants. We did not find significant differences between the genders in the correlation values, obtaining results similar to those of Figure~\ref{fig:screentime_correlation} for both male and female participants. However, slightly higher correlation values were found in different age groups. For example, in the $20$-$30$ age group, we can cite the ``Median screen time''-\textit{Activities\_Q2} pair ($r_m = 0.27, p < 0.01$), or the ``Total screen time''-\textit{Activities\_Q1} pair ($r_m = 0.29, p < 0.01$) in the $60$-$70$ age group. However, other age groups did not show these higher correlation values in any pair.

Finally, as we commented in Section~\ref{sec:data_collection}, most of the participants installed and used the mobile data collection application on the study phones provided by the project, and only $10$ participants used their private phones. We hypothesized that using the application on the study phones affected the social patterns of the participants, as it could change the way they interact with their mobile phones. Table~\ref{tab:phone_correlation} presents the strongest associations found in the group that conducted the study on their personal phones. Some other characteristic associations $-$ PRO also presented higher values in this analysis, but a \textit{p-value} above $0.1$ made us discard these as interesting associations. As we can observe, the correlation values are not as high as the ones obtained in the previously cohort-based analysis (see Table~\ref{tab:cohort_correlation}, all of them surpass the highest association found when analyzing the whole corpus of participants (i.e. $r_m = 0.11, p < 0.01$). It should be noted here that of the $10$ participants using their private phones, $9$ of them were from the RA and PD cohort groups.


\begin{table}[]
    \centering
    \tiny
    \begin{tabular}{l|c|c|c|c}
    \hline
    \textbf{Feature/PRO pair}&$\mathbf{r_m}$&$\mathbf{\rho}$&\textbf{\# points}&\textbf{\# part.}\\
    \hline\hline
    Median screen time $-$ Feel\_Q$1$ &$0.2302$&$<0.01$&$61$&$8$\\
    \hline
    Median screen time $-$ Feel\_Q$2$&$0.1685$&$0.03$&$76$&$8$\\
    \hline
    Total screen time $-$ Feel\_Q$1$&$0.1684$&$0.04$&$44$&$15$\\
    \hline
    Maximum screen time $-$ Feel\_Q$5$&$-0.1679$&$0.09$&-&-\\
    \hline
    Minimum screen time $-$ Feel\_Q$5$&$-0.1587$&$0.09$&-&-\\
    \hline
    Median screen time $-$ Woke\_Up\_Time&$-0.150$&$0.05$&$64$&$8$\\
    \hline
    
    \end{tabular}
    \caption{Top screen event feature $-$ PRO pairs associations in the group using the SMA in their personal phones.}
    \label{tab:phone_correlation}
\end{table}
\normalsize

\section{Conclusions}\label{sec:conclusions}

In this work, we explored the use of mobile data as an objective measure of fatigue, sleep, and daily activities in patients with $6$ different neurodegenerative disorders and immune-mediated inflammatory diseases. Our data comprise records reflecting how participants interact with their mobile devices, which we obtained using a mobile application. This application was also used to collect patient reported outcomes (PROs) of the symptoms mentioned above through daily questionnaires. We extracted hand-made features related to screen time and app usage information and performed statistical analysis to assess the association between mobile features and PROs through repeated measures correlation~\cite{bakdash2016repeated}.

Low correlation values between  mobile data and PROs were found in our experiments when analyzing the whole corpus. We repeated the experiments by demographic and cohort groups. Taking into account demographics, we did not find significant differences due to sex, but higher correlation values were found in some age groups. We also found higher correlation values in different cohort groups compared to the healthy control group. However, none of these values was strong enough to confirm a clear linear correlation between mobile features and PROs. Given that most patients installed the application on study phones, we discussed the possibility of finding higher correlation values when participants interact with their personal phones, as this fact could affect interaction. 

In light of the results of this work, we suggest further analysis using more complex mobile features, such as touchscreen dynamics or data from other sensors that are usually included in current smartphones. Such features have been shown to have a lot of information about smartphone interaction~\cite{2024_ESWA_SwipeFormer_Delgado}.

\section{Acknowledgment}

This research was supported by the IDEA-FAST project, which has received funding from the Innovative Medicines Initiative 2 Joint Undertaking under grant agreement No. 853981. This Joint Undertaking receives support from the European Union’s Horizon 2020 research and innovation program and EFPIA and associated partners. The work has been conducted within the ELLIS Unit Madrid.


%


%
%
%
\bibliographystyle{IEEEtran}
\bibliography{IEEEabrv,bibliography}

\end{document}